\documentclass[a4paper,11pt]{article}
\pdfoutput=1 

\usepackage{jcappub} 

\usepackage[T1]{fontenc} 

\title{\boldmath Scalar field dark matter with a cosh potential, revisited}

\author[a,1]{L. Arturo Ure\~na-L\'opez,\note{Corresponding author.}}

\affiliation[a]{Departamento de F\'isica, DCI, Campus Le\'on, Universidad de Guanajuato, 37150, Le\'on, Guanajuato, M\'exico}

\emailAdd{lurena@ugto.mx}

\abstract{Dark matter models in which the constituent particle is an ultra-light boson have become part of the mainstream discussion in cosmology and astrophysics. At the classical level, the models are represented by the dynamics of a (real or complex) scalar field endowed with a potential that contains its self-interactions, and for this reason are generically known as scalar field dark matter models. Here, we revisit the properties of such a model with a cosh potential and compare it with other known examples in the literature. Within the cosmological context, the self-interaction in the potential induces a radiation-like behavior at early times of the scalar field density, which is followed by a proper matter-like behavior at the onset of rapid field oscillations around the minimum of the potential. The solutions are found by numerical means, and from them we obtain information about the cosmological observables up to the level of linear density perturbations. We also study the general properties of self-gravitating objects in the non-relativistic limit and determine the role in them of the self-interaction obtained from the cosh potential. An overall conclusion is that, for the range of values in its parameters allowed by different constraints, a cosh potential behaves almost indistinguishable from the simpler quadratic one, which also means that the two models suffer the same tight constraints from cosmological and astrophysical observations.}

\begin{document}
\maketitle
\flushbottom

\section{Introduction} \label{sec:intro}
There is no doubt that one of the most fascinating riddles of modern
cosmology is the dark matter (DM) that seems to be an ubiquitous
component in the universe, and specially one that is indispensable for
the formation of cosmological structure. DM is an essential part of
the successful model Lambda Cold Dark Matter ($\Lambda$CDM), which has
become the standard paradigm to understand the cosmos and its
evolution up to its present state. In this model, DM is simply
described by collisionless particles that interacts mostly
gravitationally with other matter components, and makes up about
$26\%$ of the total matter budget. Quite amazingly, the theoretical
predictions of the $\Lambda$CDM model agree well with a wide range of
cosmological and astrophysical
observations~\citep{PlanckCollaboration2018,Bull2016,Bertone2016}. However,
the physical properties of the DM component remain as evasive as ever,
mostly because the particle of the standard model that describes it as Weakly
Interacting Massive Particle (WIMP), seems to be absent from most of
the so-called direct detection experiments (eg~\citep{Akerib2014,Aprile2017,Akerib2018,Baudis2018}).

Given the crisis of the WIMP hypothesis, a major trend in modern
studies about DM is characterized by the 'no stone left
unturned' approach~\citep{Bertone2018}, which asks for a thorough search of
different alternatives to the standard CDM model. One possibility that
has shown a rich phenomenology is the so-called Scalar Field Dark
Matter (SFDM) model, which has been studied relentlessly for almost
two decades now (under different names: fuzzy dark matter, wave dark
matter, ultra-light axion particles, etc),
see~\citep{Hui2017,Magana2012,Marsh2016,Lee:2017qve}. The common characteristics
in all these variants are, firstly, the presence of a scalar field (SF), whether complex or real, which is endowed with a potential that contains, explicitly or implicitly, a mass term of the form $m^2_a \phi^2$, and secondly the (bare) mass of the SF is very light, of the order of $m_a \sim 10^{-22} \, \mathrm{eV}/c^2$\footnote{Herafter, and for purposes of simplicity in the notation of the mathematical expressions, we will use natural units with $c=1=\hbar$.}. The foregoing properties are enough for the rich phenomenology we mentioned before, that allows the comparison of the model with a wide range of cosmological and astrophysical data, including, among others, gravitational waves, black holes, $21$-cm constraints, etc, see~\citep{Amendola2006,Hlozek2015a,Urena-Lopez2016,Rosa:2017ury,Ikeda2019,Barack:2018yly,Brito2017,Nebrin2018,Sarkar2016,Baumann2018,Khlopov:1985jw} for some selected examples.

The properties of SFDM can be extended if one includes the presence of
a self-interacting term of the fourth order, in the form $V(\phi) =
(1/2)m^2_a \phi^2 + (g_4/4) \phi^4$, where $g_4 > 0$ is a
dimensionless constant. Self-interacting SF models have also been
studied and their signatures as DM model have been widely discussed
in~\citep{Fan2016,Li2014,RINDLER-DALLER2014,Goodman2000}. One can even
consider the inclusion of higher order terms in the SF potential. The most famous case is the axion-like potential,
\begin{equation}
  V(\phi) = m^2_a f^2_a \left[ 1-\cos\left( \phi/f_a \right) \right] = \frac{m^2_a}{2} \phi^2 - \frac{m^2_a}{24 f^2_a} \phi^4 + \ldots \, , \label{eq:00}
\end{equation}
where $f_a$ is called, for historical reasons, the axion decay
constant. This type of models are known as axion-like particles
(ALP)~\citep{Marsh2016a}, and their anharmonic nature produces observable signatures. The most noticeable effect appears for the evolution of linear density perturbations: there is an overgrowth of the density contrast with respect to the CDM case, which is characterized as a bump in the mass power spectrum (MPS)~\citep{Zhang2017c,Zhang2017b,Cedeno2017}.

As for the non-linear formation of structure, one has to consider to the non-relativistic limit of the Einstein-Klein-Gordon (EKG) system, which is the so-called Schrodinger-Poisson (SP) system~\citep{Ruffini1969,Seidel1990,Guzman2004a,Guzman2006}. There was early evidence about the similarities between the CDM model and the solutions of the SP system~\citep{Widrow1993,Woo2009}, but it was until the results in~\citep{Schive2014}, see also~\citep{Schive2016,Mocz2017,Amin2019,Li2018}, that there appeared a clear and separate picture for the formation of structure under the SFDM hypothesis, specially for the differences with respect to CDM at small scales. 

The gravitationally bounded objects that one could identify as DM
galaxy halos all have a common structure: one central soliton surrounded by a Navarro-Frenk-White-like envelope created by the interference of the Schrodinger wave functions\citep{Schive2014}, features that have been confirmed by dedicated numerical simulations~\citep{Schwabe2016,Veltmaat2016,Veltmaat2018,Du2017,Du2018}. 

The presence of a central soliton in all of SFDM galaxy halos has
motivated studies about galactic kinematics to infer, first, the
presence of such soliton structure, and, second, to determine the mass
scale of the underlying SF
particle~\citep{Schive2014,Marsh2015a,Gonzalez-Morales2012,Gonzalez-Morales2017,Bernal2018,DeMartino2018,Calabrese2016,Marsh2018,Lora2015,Lora2012,Urena-Lopez2017,Robles2015,Robles2018,Bar2018a,Bar:2019bqz,Broadhurst2019a,Lee2019}. The
results are not yet conclusive, and depending on the analysis one may
argue for the presence of a soliton object and a SF mass of
around $m_{a 22} \equiv (m_a/10^{-22} \mathrm{eV}) \simeq 1$, or just an upper bound for the latter,
$m_{a 22} < 0.4$. More recently, the very presence of a soliton structure
has been tested using data from rotation curves in galaxies, and then
it is inferred that $m_{a22} > 10$~\citep{Bar2018a,Bar:2019bqz}.

The foregoing results on the SF mass, given that small scale structure seems to require light masses, are in tension, to say it mildly, with Lyman-$\alpha$ observations, which in contrast seem to demand larger values of the SF mass~\citep{Amendola2006,Irsic2017,Armengaud2017}. According to the latter, the lower bound on the SF mass is $m_{a 22} > 21.3$, but this result is obtained from $N$-body simulations that cannot yet capture the whole properties of the SFDM model~\citep{Nori2018AX-GADGET:Models,Nori2018a}, see for instance~\citep{Zhang2018,Zhang2017,Zhang2017d,Li2018} for critical comments.

Given the motivations above, and as an added contribution to the studies of SFDM, the main aim in this paper is to revisit the properties of the hyperbolic counterpart of the axion-like potential~\eqref{eq:00}, the cosh potential that was first studied in~\citep{Sahni2000,Matos2000,Matos2001,Matos2004},
\begin{equation}
  V(\phi) = m^2_a f^2_a \left[\cosh\left( \phi/f_a \right) - 1\right] = \frac{m^2_a}{2} \phi^2 + \frac{m^2_a}{24 f^2_a} \phi^4 + \ldots\, . \label{eq:0}
\end{equation}

Although we have made an expansion of the cosh potential~\eqref{eq:0}
similarly to that of the axion-like one~\eqref{eq:00}, there are
differences that go beyond the fourth order. Firstly, the cosh
potential resembles the exponential one $V(\phi) \simeq (m^2_a f^2/2)
e^{\pm \phi/f_a}$ for $|\phi|/f_a \gg 1$, whereas it becomes the
standard free one $V(\phi) \simeq (m^2_a/2) \phi^2$ for $|\phi|/f_a
\ll 1$. The latter is the desired form for the cosh potential to work
as CDM at late times, whereas the former is an additional advantage of
the cosh potential that has been used before to avoid any fine tuning
of the initial conditions within a cosmological
setting~\citep{Ferreira1998,Ferreira1997,Copeland1998,Sahni2000,Matos2001,Matos:2009hf}. However,
the full properties of the cosh potential~\eqref{eq:0} were left aside
and remain quite understudied (although see~\citep{Beyer2014} for a
similar model), and we intend here to cover this gap.

A summary of the paper is as follows. In Sec.~\ref{sec:cosmological-se} we revisit the cosmological solutions, in an expanding universe, of the cosh potential~\eqref{eq:0}, for both the background and linear perturbations of the SF. In doing that we will use the same numerical approach as in~\citep{Urena-Lopez2016,Cedeno2017}, based on an amended version of the Boltzmann code CLASS (Cosmic Linear Anisotropic Solving System,~\citep{Lesgourgues:2011re,Blas2011,Lesgourgues2011b,Lesgourgues2011}). We will use the numerical solutions to uncover the differences between the cosh potential and other SFDM models, in particular regarding the early evolution of the SF density and the final form of the MPS.

In Sec.~\ref{sec:self-grav}, we study the properties of the self-gravitating objects that can be formed under the cosh potential~\eqref{eq:0}, but only in the non-relativistic limit and under the quartic approximation. These approximations are justified, as our purpose is to use the numerical solutions to obtain models for small galaxy halos. We shall show that, because of the cosmological constraints, the self-gravitating objects show a different scaling symmetry with respect to the free case, and that its positive self-interaction, as opposed to the negative one for the axion-like potential~\eqref{eq:00}, has some advantages for their formation and gravitational stability. Finally, we discuss the main results and conclusions in Sec.~\ref{sec:discussion-and}.

\section{Cosmological setup \label{sec:cosmological-se}}
Here we present the relevant equations of motion of SFDM with the cosh potential~\eqref{eq:0}, for both the background and
linear perturbations. As mentiones before, we deal with the stage of
rapid oscillations following the work
in~\citep{Urena-Lopez2016,Cedeno2017}, which is based on the polar
transformation of the Klein-Gordon equation originally presented
in~\citep{Urena-Lopez2014,Reyes-Ibarra:2010jje} for inflationary
models (see also~\citep{Roy2018,Roy:2013wqa} for applications to
quintessence models of dark energy)\footnote{The equation of motion~(\ref{eq:kge}) can be solved directly, see for
  instance the study about the so-called $\alpha$-attractors
  in~\cite{Mishra:2017ehw,Cedeno:2019cgr}, where one finds a detailed
  study of the background and perturbed quantities corresponding to,
  among others, different SFDM models. It is known, however, that such
  direct approach is difficult, in numerical terms, because of the
  rapid oscillations of the SF at late times. So far, the formalism
  in~\citep{Urena-Lopez2014,Cedeno2017}, the same used here, appears
  as the most adequate to include SFDM models in Boltzmann codes.}.

\subsection{Background Dynamics \label{sec:background-dy}}
The equation of motion for a scalar field $\phi$ endowed with the potential~\eqref{eq:0}, in a homogeneous and isotropic space-time with null spatial curvature, is
\begin{equation}
        \ddot{\phi} = -3 H \dot{\phi} - m_a^2 f_a \sinh(\phi/f_a)  \, , \label{eq:kge}
\end{equation}
where a dot denotes derivative with respect to cosmic time $t$, and $H
= \dot{a}/a$ is the Hubble parameter with $a(t)$ the scale factor. The
scalar field energy density and pressure are given, respectively, by
the canonical expressions: $\rho_\phi = (1/2)\dot{\phi}^2+V(\phi)$ and
$p_\phi = (1/2)\dot{\phi}^2-V(\phi)$, whereas other matter components
are the same as in the standard cosmological model.

We define a new set of variables adapted specifically for the cosh potential,
\begin{subequations}
  \label{eq:newvars}
\begin{eqnarray}
\frac{\kappa \dot{\phi}}{\sqrt{6} H} &\equiv&  \Omega^{1/2}_\phi \sin(\theta/2) \, , \label{eq:newvarsa} \\
\frac{\kappa V^{1/2}}{\sqrt{3} H} &\equiv& \Omega^{1/2}_\phi \cos(\theta/2) = -\sqrt{\frac{2}{3}} \frac{\kappa m_a f_a}{H} \sinh(\phi/2f_a) \, , \label{eq:newvarsb} \\
y_1 &\equiv& -2 \sqrt{2} \frac{\partial_\phi V^{1/2}}{H} = 2 \frac{m_a}{H} \cosh(\phi/2f_a)\, , \label{eq:newvarsc}
\end{eqnarray}
\end{subequations}
with $\Omega_\phi = \kappa^2 \rho_\phi/3H^2$ the standard SF density
parameter, and $\kappa^2 = 8\pi G$. Similarly to the trigonometric
case~\citep{Cedeno2017}, there is a constraint equation that arises
from the known identity $\cosh^2 x - \sinh^2 x = 1$, which in terms of
the new variables~\eqref{eq:newvars} reads
\begin{equation}
    y^2_1 = 4 \frac{m^2_a}{H^2} - \lambda \Omega_\phi \left( 1+\cos \theta \right) \, , \label{eq:constraint}
\end{equation}
where for convenience we have defined the parameter $\lambda =
-3/\kappa^2 f^2$. Being an intrinsic mathematical property of the cosh
potential~\eqref{eq:0}, Eq.~\eqref{eq:constraint} should be satisfied
at all times during the cosmic evolution. 

After some straightforward algebra, the KG equation~\eqref{eq:kge} takes the form of the following dynamical system:
\begin{subequations}
\label{eq:new4}
 \begin{eqnarray}
  \theta^\prime &=& -3 \sin \theta + y_1 \, , \label{eq:new4a} \\
    y^\prime_1 &=& \frac{3}{2}\left( 1 + w_{tot} \right) y_1 + \frac{\lambda}{2} \Omega_\phi \sin \theta \, , \label{eq:new4b} \\
  \Omega^\prime_\phi &=& 3 (w_{tot} + \cos\theta)
  \Omega_\phi \label{eq:new4c} \, ,
\end{eqnarray}
\end{subequations}
Here, a prime denotes derivative with respect to the number of
$e$-foldings $N \equiv \ln (a/a_i)$, with $a$ the scale factor of the
universe and $a_i$ its initial value, and the total equation of state
(EoS) is $w_{tot} = p_{tot}/\rho_{tot}$. For the SF itself, the EoS in
Eq.~\eqref{eq:new4c}, after the proper substitution of
Eqs.~(\ref{eq:newvarsa}) and~(\ref{eq:newvarsb}), is explicitly given
by
\begin{equation}
  \label{eq:7}
  w_\phi = \frac{p_\phi}{\rho_\phi} = \frac{(1/2)\dot{\phi}^2 -
    V}{(1/2)\dot{\phi}^2 + V} = \sin^2(\theta/2) -
    \cos^2(\theta/2) = -\cos \theta \, .
\end{equation}

Notice that the dynamical system~\eqref{eq:new4} is of general applicability for the three most known examples of SFDM. In fact, for $\lambda=0$, the dynamical system becomes that of the free case~\citep{Urena-Lopez2016,Hlozek2015a,Amendola2006}, whereas for $\lambda > 0$ (which formally requires $f^2_a < 0$ so that we can change from hyperbolic to trigonometric functions) becomes that of the axion (trigonometric) case~\citep{Cedeno2017,Zhang2017c,Zhang2017b}

One critical step in the numerical solution of the equations of
motion~\eqref{eq:new4} is to find the correct initial conditions of
the dynamical variables. For that, we consider that well within
radiation domination the SF amplitude is such that $|\phi|/f \gg 1$,
and then the potential~\eqref{eq:0} is approximated by an exponential
one. Under this assumption, the field rapidly approaches the known
scaling solution of exponential
potentials~\citep{Ferreira1997,Ferreira1998,Copeland1998}, in which
the SF density evolves like the dominant background component. For the present
case, the SF must keep, initially, a constant ratio with respect to
the radiation density, $\rho_\phi/\rho_r = \mathrm{const.}$, and its
equation of state must then be $w_\phi = 1/3$. Hence, the natural
expressions for the initial conditions are,
\label{eq:match}
\begin{equation}
    \Omega_{\phi i} = -12/\lambda \, , \quad \cos\theta_i = -1/3 \, . \label{eq:scaling} 
\end{equation}

Eq.~\eqref{eq:scaling} fixes the initial values of two variables only, which leaves $y_{1i}$ as the available parameter to adjust the final contribution of $\Omega_\phi$ as the DM component. This requires to match the early exponential behavior of potential~\eqref{eq:0} to its quadratic one at late times. The precise matching is explained in~\citep{Matos2001} (see their Eq.~22), and the resultant relation between the SF mass $m_a$ and $\lambda$ is written as,
\begin{equation}
    m_{a22} \simeq 2.57 \times 10^{-10} h \, \left( \frac{\lambda}{12} \right)^2 \left( 1- \frac{12}{\lambda} \right)^{3/2} \frac{\Omega^2_{\phi 0}}{\Omega^{3/2}_{r0}} \, , \label{eq:matcha}
\end{equation}
where $h$ is the reduced Hubble parameter, and $\Omega_{r0}$ ($\Omega_{\phi 0}$) is the present value of the density parameter of radiation (SF). 

From here we can calculate an expression that can be useful for the
estimation of the initial conditions, in terms of the initial value of
the SF mass to Hubble ratio. After some straightforward algebra, we
obtain from Eq.~\eqref{eq:matcha} that
\begin{subequations}
\begin{equation}
    \frac{m_a}{H_i} = 1.5 \left[ \left( \frac{\lambda}{3} - 4 \right) \frac{\Omega_{\phi0}}{\Omega_{r0}} a_i \right]^2 \, . \label{eq:matchb}
\end{equation}

In consequence, the initial condition required for variable $y_1$ can be written, after combining Eqs.~\eqref{eq:constraint} and~\eqref{eq:scaling} together, as
\begin{equation}
    y_{1i} = \sqrt{8} \left( 1 + \frac{1}{2} \frac{m^2_a}{H^2_i} \right)^{1/2} \, . \label{eq:initialb}
\end{equation}
\end{subequations}
In practice, we calculate the initial mass to Hubble ratio in the form
$m_a/H_i = A \times$[Eq.~\eqref{eq:matchb}], where $A$ is a numerical
coefficient that is adjusted using a shooting method already within
CLASS to obtain the desired value $\Omega_{\phi 0}$ at the present
time. The system converges rapidly after a few iterations, and the
resultant SF mass to Hubble ratio is substituted into
Eq.~\eqref{eq:initialb} to complete the initial conditions\footnote{It
can be shown that the chosen initial conditions correspond to a
critical point of the dynamical system~(\ref{eq:new4}). From the
latter we find the critical conditions $y_{1c} = 3\sin \theta_c$,
$(12+ \lambda \Omega_{\phi c}) \sin \theta_c =0$ and $(w_{tot}-\cos
\theta_c) \Omega_{\phi c} =0$. It can be seen that the only critical
solution with a non-trivial density contribution is $\Omega_{\phi c} =
-12/\lambda$, $\cos \theta_c = -w_{tot}$ and $y_{1c} = 3
\sqrt{1-w^2_{tot}}$. If we take $w_{tot}=1/3$, as corresponds to
radiation domination, we recover Eqs.~(\ref{eq:scaling})
and $y_{1c} = \sqrt{8}$. Finally, Eq.~(\ref{eq:initialb}) shows the
small deviation of $y_{1i}$ from the critical point that is required
to fulfill the late time conditions in the SF evolution. See also the
Appendix~\ref{sec:init-cond-from} for the explicit expressions of the
initial conditions in terms of the original SF variables $\phi$ and $\dot{\phi}$.}.

\begin{figure*}[ht!]
\includegraphics[width=0.5\linewidth]{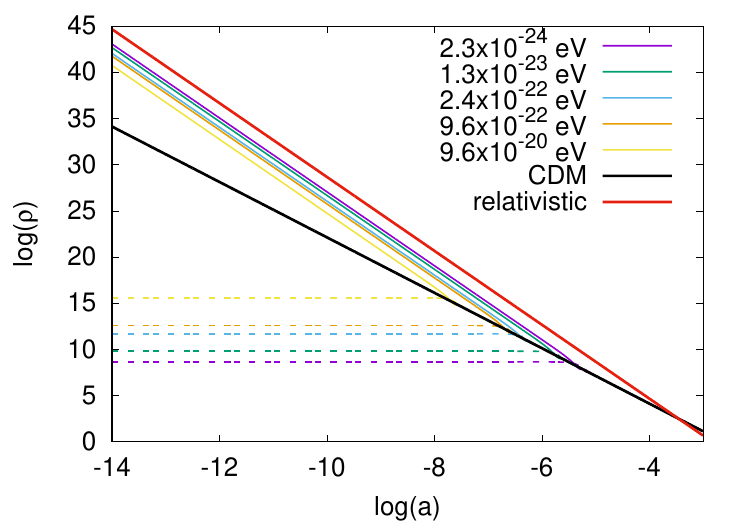}
\includegraphics[width=0.5\linewidth]{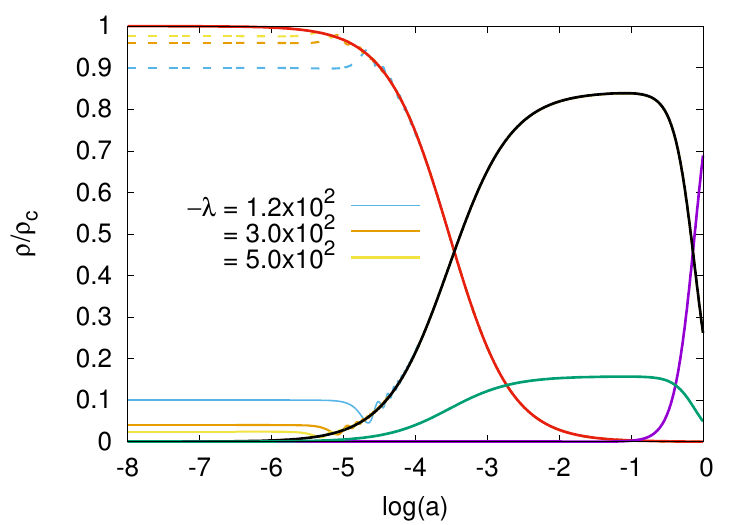}
\caption{(Left panel) The evolution of the SF and CDM energy densities as a function of the scale factor $a$, shown are the cases corresponding to different values of the SF mass $m_a$. The SF density behaves like a relativistic component (red solid line) at early times, but matches the CDM density (black solid line) after the onset of the rapid field oscillations. For comparison we also show the density of the free case with the same values of the SF mass (dashed lines with the corresponding color). (Right panel) The normalized densities, in terms of the critical one, of the different matter components: relativistic (red), CDM (black), baryons (green), and $\Lambda$ (magenta). It can be clearly seen that the SF has a non-negligible contribution as radiation at early times, which decreases for larger values of the interaction parameter $\lambda$. \label{fig:rho-omega}}
\end{figure*}

In the left panel of Fig.~\ref{fig:rho-omega} we show the evolution of the SF density $\rho_\phi$, for different values of the field mass $m_a$, in comparison with that of the standard CDM and radiation components of the fiducial $\Lambda$CDM model in~\citep{PlanckCollaboration2018}. It can be clearly seen that initially the SF density redshifts as radiation, and eventually matches the CDM line during the stage of rapid oscillations. The different curves are labeled in terms of the SF mass, to ease the comparison with other SFDM models. In Table~\ref{tab:values} we report the values of $\lambda$ and the corresponding values of the SF mass $m_a$; the latter were read out from the numerical solutions, but it can be verified that they agree very well with those estimated directly from Eq.~\eqref{eq:matcha}.

\begin{table}[htp!]
\centering
\begin{tabular}{cccc}
\hline
$-\lambda$ & $m_a$ [eV] & $\Lambda_g$ & $M_{\rm max}$ [$M_\odot$] \\ \hline
 \hline 
$1.2 \times 10^2$ & $1.1 \times 10^{-25}$ & $1.0 \times 10^1$ & $7.9 \times 10^{14}$ \\
$1.2 \times 10^3$ & $1.3 \times 10^{-23}$ & $1.0 \times 10^2$ & $2.2 \times 10^{13}$ \\
$3.6 \times 10^3$ & $1.2 \times 10^{-22}$ & $3.0 \times 10^2$ & $4.1 \times 10^{12}$ \\
$1.2 \times 10^4$ & $1.4 \times 10^{-21}$ & $1.0 \times 10^3$ & $6.8 \times 10^{11}$ \\
$1.2 \times 10^5$ & $1.4 \times 10^{-19}$ & $1.0 \times 10^4$ & $2.1 \times 10^{10}$ \\ \hline
\end{tabular}
\caption{Different values of the free parameter $\lambda$ and their
  corresponding mass values $m_a$. The latter are closed to those
  obtained from Eq.~\eqref{eq:matcha} for the fiducial values
  $\Omega_{\phi 0} = 0.26$, $\Omega_{r0} = 9.13 \times 10^{-5}$, and
  $h=0.67$. The self-interaction strength $\Lambda_g$ and the maximum
  mass $M_{\rm max}$ indicated here refer to those of self-gravitation
  objects in Eqs.~\eqref{eq:quartic} and~\eqref{eq:massesb}, respectively. \label{tab:values}}
\end{table}

We see that the matching of SFDM to CDM happens later (earlier) for smaller (larger) values of the SF mass, which is a common feature of SFDM models. For a quick comparison, we also plotted the results from the free case (dashed lines), which shows that both models start their rapid oscillations at the same time. This is not surprising, as the evolution of the SFDM with a cosh potential can be parametrized in terms of the SF mass only (recall Eq.~\eqref{eq:matchb}), so that the appearance of the rapid oscillations is solely dictated by the mass to Hubble ratio $m_a/H$, and then its late behavior should match that of the free model too.

As mentioned before, the SF density has a non-negligible contribution during radiation domination, but we note that such contribution diminishes for larger values of the SF mass. For a better illustration of this effect, we show the ratio of the different matter components with respect to the critical density in the right panel of Fig.~\ref{fig:rho-omega} (see also~\citep{Matos2000,Matos2001}), and make a comparison again with the results from $\Lambda$CDM. In this form, the contribution of the SF component is noticeable during radiation domination. Also, it must be noticed that the radiation, matter, $\Lambda$ domination eras do not suffer any alteration and proceed just like in the $\Lambda$CDM case.

\begin{figure}[ht!]
\centering
\includegraphics[width=0.5\linewidth]{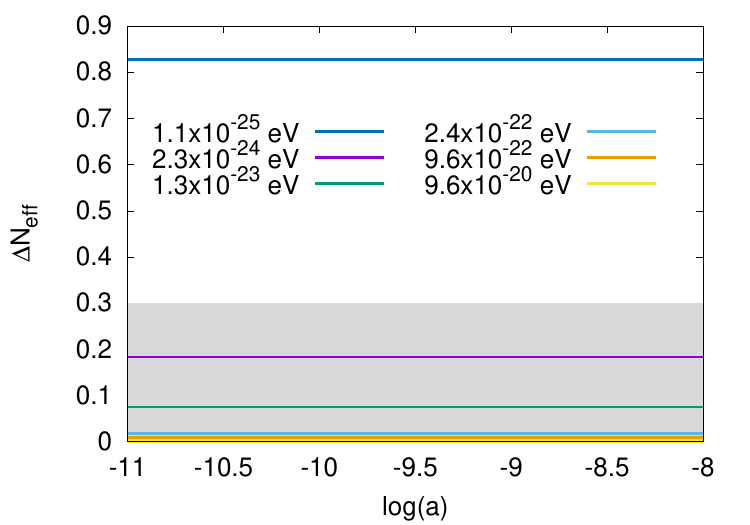}
\caption{The extra relativistic degrees of freedom $\Delta N_{\rm eff}$ induced by the non-negligible contribution of the SF density at around the time of nucleosynthesis. The shaded region represents the reported constraint in~\citep{PlanckCollaboration2018}, $\Delta N_{\rm eff} < 0.3$ at $95\%$CL. \label{fig:nucleo}}
\end{figure}

One common concern about the presence of the SF at early times is its
contribution as an extra relativistic degree of freedom, that
increases the expansion rate at the time of nucleosynthesis. The
relativistic degrees of freedom, when the SF density at early times is
included, are given by $N_{\rm eff} = (\rho_{\rm rad} + \rho_\phi)/\rho_\nu$, with $\rho_\nu$ the neutrino density. The resultant effect is shown in Fig.~\ref{fig:nucleo}, for the indicated values of the SF mass and in terms of the quantity $\Delta N_{\rm eff} \equiv N_{\rm eff} -3.046$. 

The non-negligible contribution of the SF to $\Delta N_{\rm eff}$ can be clearly seen during radiation domination and at the time around nucleosynthesis (see also Fig.~\ref{fig:rho-omega}). The present constraints, from~\citep{PlanckCollaboration2018}, then indicate that $\Omega_\phi < 3.5 \times 10^{-2}$, which in turn translates, by means of Eq.~\eqref{eq:scaling}, into $-\lambda > 3.4 \times 10^2$. Considering the values shown in Table~\ref{tab:values}, the cosh potential~\eqref{eq:0} then evades the nucleosynthesis constraints as long as its field mass is $m_{a22} > 10^{-2} $.

As a final note, we contrast the nucleosynthesis constraint on the
cosh potential to that applied on a SFDM with a quartic
self-interaction (although with a complex SF)~\citep{Li2014}: for the
latter there is a kinetic dominate phase (which manifests as a
stiff-fluid behavior) prior to the radiation-like behavior of the
SF. Thus, one has to finely tune the values of the SF mass and the
quartic interacting parameter to avoid any undesirable effects upon
the nucleosynthesis process\footnote{A similar study of the quartic
  self-interaction model for a real scalar field can be found in~\citep{Cembranos2018},
  although the initial conditions were seemingly chosen so that the
  early stiff and radiation like behaviors were avoided. In such a case,
  the nucleosynthesis constraints are not important, and both the mass
  and self-interaction parameters can be varied more freely than in
  the case considered in~\citep{Li2014}.}. The cosh
potential~\eqref{eq:0} evades such tight constraints because the early
radiation-like solution~\eqref{eq:scaling} is a quite stable attractor solution of the equations of motion~\eqref{eq:new4}.

\subsection{Linear Density Perturbations \label{sec:linear-den}} 
We now turn our attention to the linear field perturbations $\varphi$ around the background value in the form $\phi(x,t) = \phi(t)+\varphi(x,t)$. As in previous works, we choose the synchronous gauge with the line element $ds^2 = -dt^2+a^2(t)(\delta_{ij}+\bar{h}_{ij})dx^idx^j$, where $\bar{h}_{ij}$ is the tensor of metric perturbations. The equation of motion for a given Fourier mode $\varphi(k,t)$ reads \citep{Ratra1988,Ferreira1997,Ferreira1998,Perrota1999}
\begin{equation}
  \ddot{\varphi} = - 3H \dot{\varphi} - \left[\frac{k^2}{a^2} +m_a^2\cosh(\phi/f)\right] \varphi -
  \frac{1}{2} \dot{\phi} \dot{\bar{h}} \, , \label{eq:13}
\end{equation}
where a dot means derivative with respect the cosmic time, $\bar{h} = {\bar{h}^j}_j$ and $k$ is a comoving wavenumber.

Following~\citep{Cedeno2017,Urena-Lopez2016}, we can choose appropriate variables to transform Eq.~\eqref{eq:13} into the following dynamical system,
\begin{subequations}
\label{eq:dp}
\begin{eqnarray}
\delta^\prime_0 &=&  \left[-3\sin\theta-\frac{k^2}{k^2_J}(1 - \cos \theta) \right] \delta_1 + \frac{k^2}{k^2_J} \sin \theta \, \delta_0 - \frac{\bar{h}^\prime}{2}(1-\cos\theta) \, , \label{eq:dp0} \\
\delta^\prime_1 &=& \left[-3\cos \theta - \left( \frac{k^2}{k^2_J} - \frac{\lambda \Omega_\phi}{2y_1} \right) \sin\theta \right] \delta_1 + \left(\frac{k^2}{k^2_J} - \frac{\lambda \Omega_\phi}{2y_1} \right) \left(1 + \cos \theta \right) \, \delta_0 - \frac{\bar{h}^\prime}{2} \sin \theta \, , \label{eq:dp1}
\end{eqnarray}
\end{subequations}
where $k_J^2 \equiv a^2 H^2 y_1$ is the (squared) Jeans wavenumber and
a prime again denotes derivative with respect to the number of
$e$-folds $N$. In this approach $\delta_0 \equiv \delta \rho_\phi/\rho_\phi$ is the SF density contrast and $\delta_1$ is a second density constrast that arises naturally under the new variables. As before for Eqs.~\eqref{eq:new4}, Eqs.~\eqref{eq:dp} have again the same structure for the cosh, axion and free potentials, and the only difference is the value of the interaction parameter $\lambda$ for each case. 

\begin{figure*}[htp!]
\includegraphics[width=0.5\linewidth]{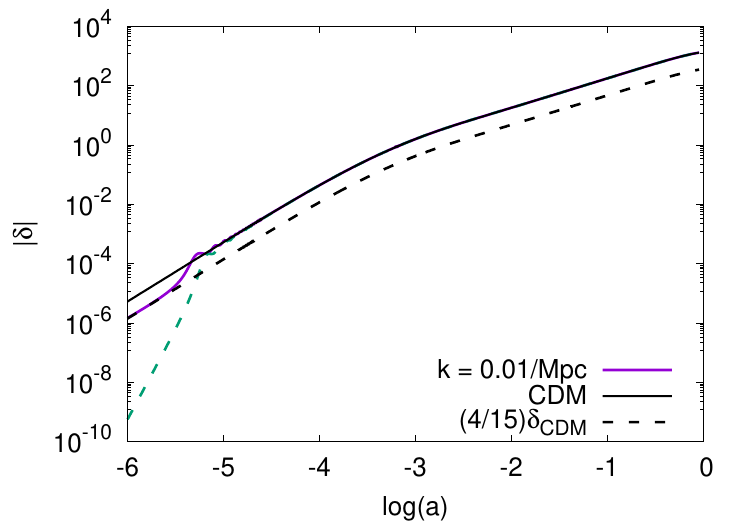}
\includegraphics[width=0.5\linewidth]{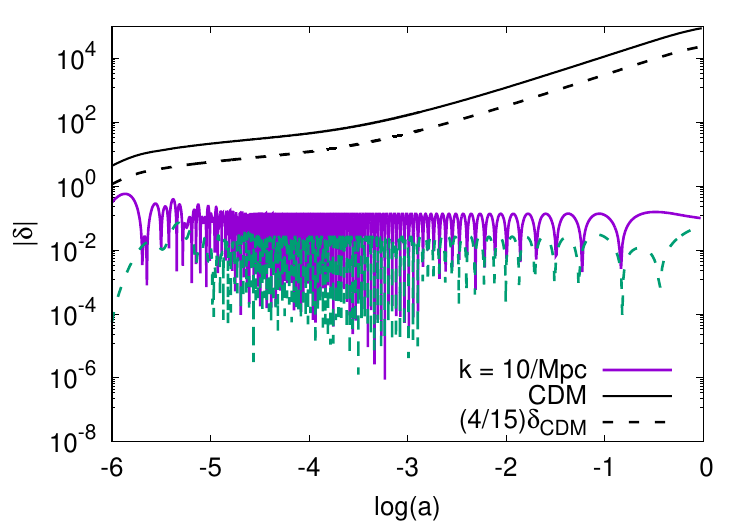}
\caption{The amplitude of the density contrast $|\delta_0|$, for
  two representative cases: large scales (left panel), and small scales
  (right panel). The SF mass in these examples is $m_{a22} = 2.3 \times
  10^{-2}$. On large scales, the SF density contrast follows first the
  attractor solution $(4/15)\delta_{\rm CDM}$ (dashed black curve),
  whereas it matches CDM exactly after the onset of rapid
  oscillations. The attractor solution is also followed at early times
  in the case of small scales, but the density contrast do not grow
  afterwards. For comparison, we also show the results corresponding
  to the free case~\citep{Urena-Lopez2016} (dashed curves in color). \label{fig:deltas}}
\end{figure*}

For the initial conditions, we use the same solutions at early times
considered for a quadratic potential in Ref.\cite{Urena-Lopez2016},
see also\cite{Cedeno2017}, which allows us to use one single Boltzmann
code for three different potentials. It must be noticed that, for the
particular case of the exponential potential, there is an attractor
solution for linear perturbations at large scales, which is $\delta_0
= (4/15) \delta_{CDM}$\citep{Matos2001,Ferreira1998}. Although we are
not considering it in the Boltzmann code, such attractor solution is
quickly reached just after some $e$-folds (see Fig.~\ref{fig:deltas}
below and the Appendix~\ref{sec:init-cond-line}).

The numerical results illustrating both the cosh and quadratic cases, for $m_{a22} = 2.3 \times 10^{-2}$ and $-\lambda = 5\times 10^2$, are shown in Fig.~\ref{fig:deltas}. The plots replicate those presented in~\citep{Matos2000,Matos2001}, which were the first to show the evolution of density perturbations for the cosh potential obtained directly from a Boltzmann code (the now outdated CMBFAST).

In the case of large enough scales, we see that the SF density contrast reaches the attractor solution at early times mentioned above, but quickly joins the standard CDM evolution once the SF oscillations start. In contrast, for small enough scales, again the density contrast reaches the same attractor solution at early times but it ends up oscillating at late times with almost a constant amplitude. The results from the cosh potential match those of the quadratic one at late times, which is a common feature in our numerical results: as long as the SF mass is the same, the two models are almost indistinguishable.

\begin{figure}[tp!]
\centering
\includegraphics[width=0.5\linewidth]{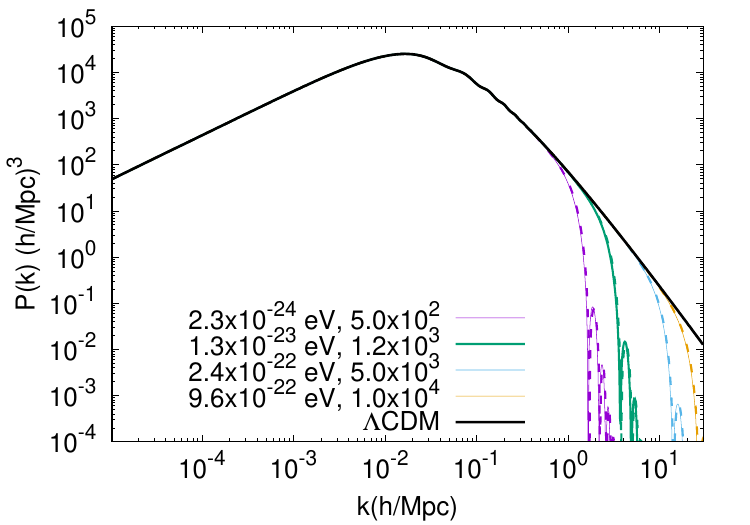}
\caption{MPS at the present time for SFDM with a cosh potential, for the values of the SF mass and $(-\lambda)$ as indicated in the plot. Also shown is the quadratic potential with the same values of the SF mass (dashed lines in colour). The MPS appears to be the same for both models, see the text for more details. \label{fig:mps}}
\end{figure}

Finally, we show in Fig.~\ref{fig:mps} the MPS at the present time for
a cosh potential, also in comparison with the results that are
obtained for the quadratic one with the same values of the SF mass. As
expected from the discussion above about the density contrast, the MPS
is practically indistinguishable in the two cases, which means that
the well-known cut-off in the MPS of the cosh potential (first
reported in~\citep{Matos2001} for any SFDM model) just depends upon
the values of the SF mass. The cosh potential is then an example in
which a more involved SF potential does not provide, within the
cosmological context, for both the background and linear
perturbations, late-time results different from those of the free case. The reason for this is, of course, the close relation between the SF mass $m_a$ and the self-interaction parameter $\lambda$ in Eq.~\eqref{eq:masses}, which is required to obtain an appropriate radiation to matter transition for a corrrect contribution of SFDM at late times.

\section{Self-gravitating objects \label{sec:self-grav}}
A separate question of physical interest is the type of self-gravitating objects that are formed under a cosh potential. This topic has been treated before in~\citep{Urena-Lopez2002b}, and here we just give a brief account of the main results reported throughout the specialized literature (see~\citep{Liebling2017,Schunck2003GeneralStars} for recent reviews). 

The key parameter here is the quartic coupling that arises from the series expansion of potential~\eqref{eq:0} up to the fourth order: $V(\phi) \simeq (m^2_a/2) \phi^2 + (m^2_a f^{-2}_a/24) \phi^4$. Following standard nomenclature~\citep{Liebling2017}, although adapted for real scalar fields (see Eqs.~\eqref{eq:GPP} below), the self-interaction strength that is of physical interest is quantified by the combination\footnote{Notice that the interaction parameter in the case of boson stars, as originally defined in~\citep{Colpi1986}, is $\Lambda_g \equiv g_4/(4\pi G m^2_a)$, where $g_4$ is the quartic coupling constant as in $(g_4/4)\phi^4$. In the case of real scalar fields, a more appropriate definition is $\Lambda_g \equiv 3 g_4/(8\pi G m^2_a)$~\citep{Urena-Lopez2002b,Urena-Lopez2012}. Once we consider the Newtonian limit of the Klein-Gordon equation for real scalar fields, it appears convenient to further take $\Lambda_g \equiv 3 g_4/(16\pi G m^2_a)$, see~\citep{Lee1996,Guzman2004a,Guzman2006}. The latter is the definition used to write $\Lambda_g$ in Eq.~\eqref{eq:quartic}.} 
\begin{equation}
\Lambda_g \equiv m^2_a f^{-2}_a/(32 \pi G m^2_a) = 3/(12\kappa^2 f^2_a) = - \lambda/12 \, , \label{eq:quartic}
\end{equation}
and then determined solely by the interaction parameter $f_a$ in the potential~\eqref{eq:0}. In passing by, notice also that the initial condition~\eqref{eq:scaling} can also be written as $\Omega_{\phi i} = \Lambda^{-1}_g$.

For the values of $\lambda$ suggested by the cosmological constraints discussed above, we find that the self-interaction strength must be of the order of $\Lambda_g \sim 10^2$ or larger. In any case, we find that cosmological constraints seem to suggest that self-gravitating objects belong to the so-called strong-regime, $\Lambda_g \gg 1$. In what follows we will revise the properties of such objects.

\subsection{General stability and equilibrium configurations \label{sec:general-stability}}
To begin with, there is a maximum mass for stable configurations
(see~\citep{Colpi1986,Balakrishna1998,Urena-Lopez2002b}), that is
approximately given by $M_{\rm max} \simeq 0.22 \Lambda^{1/2}_g
m^2_{\rm Pl}/m_a$, and which can be a very large number for the
field masses of cosmological interest\footnote{The corresponding
  maximum mass in the non-interacting case ($\Lambda_g =0$) is $M_{\rm
    max} \simeq 0.6 m^2_{\rm Pl}/m_a$, see for
  instance~\citep{Seidel1990,Seidel1991,Alcubierre2003Numerical2-oscillatons}. As
  for ALP models with the axion potential~\eqref{eq:00}, that has a
  negative self-interaction strength $\Lambda_g < 0$, the estimated
  maximum mass is $M_{\rm max} \simeq |\Lambda_g|^{-1/2} m^2_{\rm
    Pl}/m_a$~\citep{Guzman2006,Chavanis2016,Liebling2017,Helfer:2016ljl};
  then, the larger the self-interaction the less massive are stable
  configurations.}. Moreover, as we have seen in
Sec.~\ref{sec:cosmological-se} above, for the particular case of the
cosh potential~\eqref{eq:0} there exists a close relation between the
SF mass and the self-interaction strength, see Eq.~\eqref{eq:matcha},
and then the maximum mass can be written in terms of $\Lambda_g$
alone. The two aforementioned relations, for the fiducial cosmological
values assumed in Sec.~\ref{sec:cosmological-se}, are
\begin{subequations}
\label{eq:masses}
\begin{eqnarray}
    m_{a22} &\simeq& 1.37 \times 10^{-5} \, \Lambda^2_g \left( 1- 1/\Lambda_g \right)^{3/2} \, , \label{eq:massesa} \\ 
    M_{\rm max} &\simeq& 2.15 \times 10^{16} \, M_\odot \, \left( \Lambda_g - 1 \right)^{-3/2} \, . \label{eq:massesb}
\end{eqnarray}
\end{subequations}

Some examples of $M_{\rm max}$ are shown in Table~\ref{tab:values} for reference. Notice that the maximum mass could be as low as $10^{10} \, M_\odot$, which means that, for large enough values of the interaction parameter $\Lambda_g$, some configurations of astrophysical interest could be close to the point of gravitational instability. In the least troublesome case, unstable configuration just migrate to stable ones by means of gravitational cooling, but in other more involved cases they could collapse into black holes~\citep{Seidel1991,Balakrishna1998,Alcubierre2003Numerical2-oscillatons,Guzman2006,Urena-Lopez2012,Liebling2017}.

However, the case of more astrophysical interest is the
non-relativistic one which is obtained in the weak field limit of the
Einstein-Klein-Gordon (EKG) system. Under the ansatz $\sqrt{8\pi G}
\phi = e^{-im_at} \psi + \mathrm{c.c.}$, where $\psi$ is a
non-relativistic, complex wave function, the EKG system can be recast
as the Gross-Pitaevskii-Poisson (GPP) system of equations
(eg~\citep{Chavanis2011}). Using the field mass to define
dimensionless variables for time and distance, $t \to t/m_a$ and $\mathbf{x} \to \mathbf{x}/m_a$, the final form of the GPP equations is (for details see~\citep{Guzman2004a,Lee1996,Chavanis2011}),
\begin{subequations}
\label{eq:GPP}
\begin{eqnarray}
 i \dot{\psi} &=& - \frac{1}{2} \nabla^2 \psi + \left( U + \Lambda_g |\psi|^2 \right) \psi \, , \label{eq:GPPa} \\
 \nabla^2 U &=& |\psi|^2 \, , \label{eq:GPPb}
\end{eqnarray}
\end{subequations}
where $U$ is the Newtonian gravitational potential. If $\Lambda_g =0$, we recover from Eqs.~\eqref{eq:GPP} the well-known Schrodinger-Poisson (SP) system that has been the standard workhorse to study the formation of large scale structure in SFDM models (eg~\citep{Schive2014,Hwang2009,Veltmaat2016,Veltmaat2018,Mocz2017}).

The GPP system~\eqref{eq:GPP} is invariant under the following scaling symmetry,
\begin{equation}
    \left\{ t,\mathbf{x}, \psi,U, \Lambda_g \right\} \; \to \; \left\{ \alpha^{-2} \hat{t}, \alpha^{-1} \hat{\mathbf{x}}, \alpha^2 \hat{\psi}, \alpha^2 \hat{U}, \alpha^{-2} \hat{\Lambda}_g \right\} \, , \label{eq:scaling-sym}
\end{equation}
where $\alpha$ is an arbitrary parameter. This somehow eases the
numerical effort to find the equilibrium configurations of the GPP
system if we can choose an appropriate value for the scaling parameter
$\alpha$. The usual choice in the non-interacting case ($\Lambda_g
=0$) is $\alpha = |\psi_c|^{-1/2}$, so that the rescaled wavefunction
can simply have a central value of order unity, $\hat{\psi}_c
=1$~\citep{Guzman2004a}. This cannot be done anymore in the
interacting case, and then the most sensible choice is to use instead
the strength parameter to rescale the solutions, so that $\alpha =
\Lambda^{-1/2}_g$, which is equivalent to set $ \hat{\Lambda}_g
=1$. After this, there is not any further scaling available for
Eqs.~\eqref{eq:GPP} to ease the numerical analysis, and then one has
but to consider solutions case by case for different central values of
the wavefunction $\hat{\psi}_c$.

Of particular interest here are the so-called equilibrium configurations of the GPP system, which are stationary, spherically symmetric, solutions of the wave function in the form $\psi(t,r) = \varphi(r) e^{-i\sigma t}$, where $\sigma$ is a constant parameter. Upon substitution of the foregoing ansatz in Eqs.~\eqref{eq:GPP}, the resultant system of equations is
\begin{subequations}
\label{eq:GPP-stationary}
\begin{eqnarray}
 \partial_{rr} (r \varphi) &=& 2 \left( U - \sigma + \varphi^2 \right) \varphi \, , \label{eq:GPP-stationarya} \\
 \partial_{rr} (r U) &=& \varphi^2 \, . \label{eq:GPP-stationaryb}
\end{eqnarray}
\end{subequations}
Notice that we have implicitly made use of the scaling
symmetry~\eqref{eq:scaling-sym} and then the re-scaled
self-interaction parameter has been set to unity, $\hat{\Lambda}_g
=1$. For purposes of simplicity in the notation, all quantities in
Eqs.~\eqref{eq:GPP-stationary} are assumed to have been re-scaled
according to~\eqref{eq:scaling-sym}. The numerical solutions were
found by means of a standard shooting procedure to determine, for a
given value of $\varphi(0)$, the corresponding ones of $U(0)$ and
$\sigma$.

The numerical values of some of the quantities calculated
for each one of the equilibrium configurations are shown in
Table~\ref{tab:equilibrium}; for comparison, we also show the
corresponding values of the non-interacting case (which coincide with
those reported in\cite{Guzman2004a}). In general, we see that
interacting configurations are more massive and a bit larger in size than
non-interacting ones, as one can see from the comparison between the
cases with $\varphi(0)=1$. A more detailed description of differences
and similarities is given in the next section below.

\begin{table}[htp!]
\centering
\begin{tabular}{|ccccc|}
\hline
$\Lambda_g$ & $\varphi_c$ & $\sigma$ & $M_T$ &$r_{95}$ \\ \hline
0 & 1.0 & -0.69 & 2.06 & 3.93 \\
\hline
1 & 0.1 & -0.06 & 0.63 & 10.89 \\
  & 0.5 & -0.46 & 2.23 & 5.83 \\
  & 1.0 & -1.25 & 4.84 & 4.36 \\
  & 1.5 & -2.47 & 8.78 & 3.78 \\ \hline
\end{tabular}
\caption{Characteristic values of non-interacting and interacting equilibrium configurations. The quantities shown are the strength parameter $\Lambda_g$, the central field $\varphi_c$, the eigen-frequency $\sigma$, the total mass $M_T$, and the radius containing $95\%$ of the total mass $r_{95}$. \label{tab:equilibrium}}
\end{table}

\subsection{Weak and strong field limits}
It is known that there is a universal density profile for
non-interacting equilibrium configurations, which can be approximated
by the formula: $\rho(r) =
\rho_c/(1+r^2/r^2_s)^8$~\citep{Marsh2015a,Schive2014}, and whose
central density $\rho_c$ and scale radius $r_s$ are
related in the scale-independent form $\rho_c r^4_s = \mathrm{const.}$
(see Eqs.~\eqref{eq:scaling-sym}). The aforementioned density profile
has been widely used in studies of SFDM and in its comparison with diverse astrophysical data~\citep{Gonzalez-Morales2017,Herrera-Martin2019,Urena-Lopez2017,Marsh2015a,Bernal2018}.

\begin{figure*}[htp!]
\includegraphics[width=0.5\linewidth]{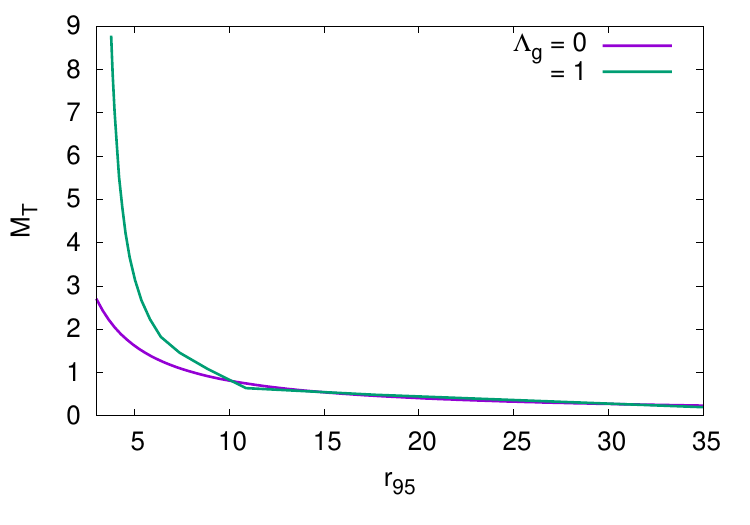}
\includegraphics[width=0.5\linewidth]{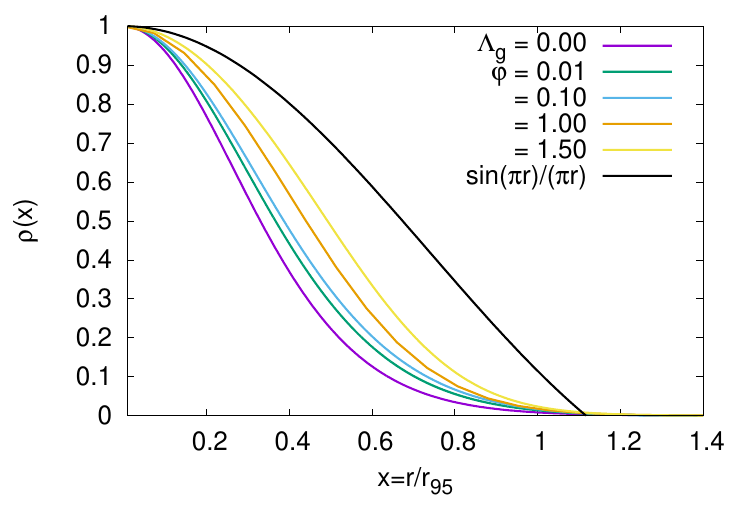}
\caption{(Left panel) The dependence of the total (numerical) mass as a function of $r_{95}$, for both the interacting and non-interacting cases. The two curves coincide for systems with a low total mass and large radii. (Right panel) Density profiles, normalized to their central values, obtained from the solution of the rescaled GPP system~\eqref{eq:GPP-stationary}, see also Eq.~\eqref{eq:GPP}. The interacting configurations have a profile that lies in between that of the non-interacting case and the analytic one $\sin(\pi r)/(\pi r)$. \label{fig:equilibriuma}}
\end{figure*}

Things are not quite straightforward in the interacting case, as one
cannot find, in general, a universal density profile that is just
rescaled for all possible configurations. However, if the central
field value is weak enough, $\hat{\varphi}_c \ll 1$, the
self-interaction term in Eq.~\eqref{eq:GPPa} must become negligible,
and in this limit we should somehow recover the non-interacting
case. This is exactly what we found in our numerical solutions, which
is also shown in the left panel of Fig.~\ref{fig:equilibriuma} for the
relation between the total mass $\hat{M}_T$ and the $95\%$ radius
$\hat{r}_{95}$. Weaker configurations have larger radius, and it is
for them that we see the equivalence between the interacting and
non-interacting cases; in terms of the central field, the equivalence
is reached for $\hat{\varphi}_c \lesssim 0.1$. The main advantage of
the equivalence between the two type of configurations is that we can
use, for the interacting case, all the known scaling symmetries of the
SP system, avoiding all the hassle of calculating equilibrium
configurations case by case.

On the other hand, there is the belief that interacting configurations
have an analytic expression for their density in the form $\rho(r) =
\rho_c \sin(\pi r/r_s)/(\pi r/r_s)$, for $r \leq r_s$ and zero
otherwise. The scale radius of the configuration is given by $r_s =
(\pi/\sqrt{2}) |\Lambda_g|^{1/2} m^{-1}_\phi$, whereas the central
density would be a free
parameter~\citep{Diez-Tejedor2014b,Chavanis2012,Lee1996}. We have not
found any evidence of such common profile in the numerical results of
the GPP system~\eqref{eq:GPP}. See, for instance, the right panel of
Fig.~\ref{fig:equilibriuma}, where we show the normalized density
profiles, $\rho(r)/\rho_c$, for some of the numerical solutions listed
in Table~\ref{tab:equilibrium}. The profiles differ one from each
other, as they correspond to different values of $\varphi_c$, and they
all differ from the analytic profile $\sin(\pi r/r_s)/(\pi r/r_s)$.

The belief in the analytic expression seems to have been originated
from the Newtonian approximation used in~\citep{Lee1996} of the
relativistic equilibrium configurations obtained in~\citep{Colpi1986}
for the limit $\Lambda_g \to \infty$. It was assumed
in~\citep{Lee1996} that the Newtonian limit of the fundamental frequency
is $\omega/m \to 1$, and then the density must obey the equation
$\nabla^2 \rho = - \rho$. However, the correct result is $\omega/m \to
1 - \sigma/2$, where $\sigma$ is the eigen-frequency in
Eq.~\eqref{eq:GPP-stationarya}. Hence, the equilibrium configuration
must be obtained again from an eigenvalue problem to determine the
correct value of $\sigma$. Actually, one can see in Fig.~3
of~\citep{Colpi1986} that the analytic profile is just an
approximation and that the actual, numerically obtained, field profile
deviates considerably from it at large values of $r$.~\footnote{The
  assumed profile with a large self-interaction, $\sin(\pi r/r_s)/(\pi
  r/r_s)$, also appears from the so-called Thomas-Fermi (TF)
  approximation, in which one neglects the spatial derivatives of the
  wavefunction $\psi$ in Eq.~(\ref{eq:GPPa}). The TF profile has
  been shown to be at variance with galactic
  observations~\citep{Arbey2003,Bohmer2007,Diez-Tejedor2014b}. It
  would be, for the reasons explained in the text, incorrect to use those negative results against SFDM models.}

As a final note in the strong field limit, we must recall that there
is a critical mass for equilibrium configurations, see
Eq.~\eqref{eq:massesb}, which also corresponds to a critical field
value, estimated to be $\kappa_0 \phi \simeq
\Lambda^{-1/2}_g$~\citep{Urena-Lopez2002b,Colpi1986}. For
non-relativistic configurations this translates, in terms of re-scaled
quantities, into the constraint $\hat{\varphi}_c \lesssim 1$. As shown
in the right panel of Fig.~\ref{fig:equilibriuma}, it may be possible
to reach the analytic profile $\sin (\pi r)(\pi r)$ if $\hat{\varphi} \gg 1$, but this means that the resultant configuration is well within the unstable branch of relativistic equilibrium solutions (as shown in~\citep{Guzman2013}), which for that very reason makes it of less interest for cosmological and astrophysical applications. 

\section{Discussion and conclusions \label{sec:discussion-and}}
We have revisited the general properties of one SFDM model with a cosh potential, which was proposed as an alternative model to CDM almost two decades ago. This is an example, similar to the axion case, in which one can include non-linear terms in the SF potential that open the door for distinguishable features with respect to the simple quadratic case, which has been used as the paradigm for SFDM models.

At the cosmological level, we reviewed the solutions of the cosh model for physical quantities in the background and their linear perturbations. The distinctive feature of the cosh model is its scaling solution during the epoch of radiation domination, under which the SF has a non-negligible contribution to the radiation component. Such scaling solution is a strong attractor solution, which helps the SFDM model to evade the fine tuning problem, to get the right DM contribution at the present time, that is unavoidable for the quadratic and axion cases. By means of the nucleosynthesis constraint for extra, non-thermal, relativistic degrees of freedom, we found an upper bound on the interaction parameter, $\kappa f_a \lesssim 9.4 \times 10^{-2}$ ($f_a \lesssim 1.9 \times 10^{-2} m_{\rm Pl}$), which is equivalent to a lower bound on the SF mass, $m_{a22} > 10^{-2}$.

As for the analysis of the linear density perturbations, we confirmed the existence of an attractor solution for them at early times, which also alleviates the fine tuning problem of SF linear perturbations. The resultant MPS summarizes well the general properties of the density perturbations with the presence of a sharp cut-off of power at small scales. Although the cut-off had already been shown to exist in~\citep{Matos2001}, we obtained here that the cut-off in the cosh case is quite similar to that of the quadratic one, and then for both cases the properties of the density perturbations are characterized by the SF mass only.

We also revisited the properties of the self-gravitating objects obtained from the cosh potential, although only for the Newtonian limit and using a quartic approximation to the potential. The self-interaction term for the cosh potential is positive definite, which then contributes with an extra repulsive force to prevent the gravitational collapse of the SF equilibrium configurations. This reinforces their gravitational stability, as the maximum mass they can attain is larger than for the non-interacting case. 

However, we made estimations for the SF amplitude required to represent realistic galaxies by means of soliton structures and found that, in such limit, the self-interaction term can be neglected. Hence, the soliton configurations from SFDM with a cosh potential can be safely represented by the standard free ones, whose known properties are closely linked to the SF mass only~\citep{Schive2014}.

Another question that usually arises within the context of SF DM models is the possible formation of a Bose-Einstein condensate (BEC). Much attention has been put forward in the formation of a BEC for the axion case ever since the results reported in~\citep{Sikivie2009}. However, such results have been heavily contested because of the attractive (negative) self-interaction attributed to axion-like potentials~\citep{Guth2015}. Actually, previous works had already shown that BEC formation is possible for repulsive (positive) self-interactions by means of an inverse particle cascade that populates the zero-mode in the distribution function of particles~\citep{PineiroOrioli2015}, whereas the opposite happens, eg the depopulation of the zero-mode by a direct particle cascade, for attractive (negative) self-interactions. The cosh potential~\eqref{eq:0} has a repulsive self-interaction, and then the formation of a BEC is an expected result that would clearly differentiate it from axion-like potentials.

A SFDM model with a cosh potential is a candidate as good as the free model to be an alternative to CDM, with the added advantages of being a non-linear potential with a positive definite self-interaction. As is currently known for such kind of DM models, there is a good agreement with observations from large (CMB, MPS) to small scales (eg dwarf spheroidals), but the question remains about its appropriateness for the full range of scales. Of particular interest are constraints from Lyman-$\alpha$ observations, which have since the study in~\citep{Amendola2006} been a point of major concern for the preferred model with $m_{a22} \sim 1$~\citep{Zhang2017b,Zhang2017d,Irsic2017,Armengaud2017,Nori2018a,Li2018}, but we may have to wait for the development of better numerical simulations before we can decide firmly about the viability of the model.

\acknowledgments
I am grateful to Varun Sahni for useful comments and suggestions about
the manuscript. This work was partially supported by Programa para el
Desarrollo Profesional Docente; Direcci\'on de Apoyo a la
Investigaci\'on y al Posgrado, Universidad de Guanajuato 010/2019; and
CONACyT M\'exico under Grants No. A1-S-17899 and 286897.

\appendix
\section{Initial conditions from the field perspective for the
  background evolution \label{sec:init-cond-from}}
Rapidly oscillating scalar fields are well-known for its late time
behavior as pressureless matter. Although there exist some analytic
approximations to describe them in such regime, the comparison with
observational data still requires numerical solutions of the SF
equations of motion. As we have discussed in Sec.~\ref{eq:match}, it
is necessary to use precise formulas to link the initial conditions of
the field variables to the wished values of the physical variables at
late times, so that the search of physically relevant solutions can be
made in a continous and systematic manner.

The use of polar variables to transform the KG of SFDM has shown to be
a fruitful method to find appropriate initial conditions of the SF
variables. As a complement to those derived in
Sec.~\ref{sec:background-dy}, here we rewrite some of the formulas so
that can be used for the standard field approach for both the cosh and
quadratic potentials.

\subsection{Cosh potential}
We first combine Eqs.~(\ref{eq:newvars}) to find the expressions of
the SF variables in terms of the polar ones,
\begin{equation}
  \label{eq:1}
  \tanh(\phi/2f_a) = - \sqrt{-2\lambda} \, \Omega^{1/2}_\phi
  \cos(\theta/2) \, y^{-1}_1 \, , \quad \kappa \dot{\phi} = \sqrt{6} H
  \Omega^{1/2}_\phi \sin(\theta/2) \, .
\end{equation}
Using the expressions for the initial conditions,
Eqs.~(\ref{eq:scaling}) and~(\ref{eq:initialb}),
we obtain,
\begin{equation}
  \label{eq:2}
   \tanh(\phi_i/2f_a) = - \left( 1 + \frac{m^2_a}{2H^2_i}
   \right)^{-1/2} \, , \quad  \kappa \dot{\phi}_i = 4 \kappa m_a f_a
   \left( \frac{m_a}{H_i} \right)^{-1} \, .
 \end{equation}
 Finally, the initial values of the SF and its time derivative are
 calculated once the matching expression~(\ref{eq:matchb}) is
 substituted in Eqs.~(\ref{eq:1}), the latter providing the link
 between the initial values and the chosen ones $\Omega_{\phi 0}$ and
 $a_i$. In general, one expects $m_a/H_i \sim a^2_i \ll 1$, and then Eqs.~(\ref{eq:2}) can be approximated as
 \begin{equation}
   \label{eq:3}
   \tanh(\phi_i/2f_a) \simeq - 1 + \frac{9}{16} \left[ \left(
       \frac{\lambda}{3} - 4 \right)
     \frac{\Omega_{\phi0}}{\Omega_{r0}} a_i \right]^4 \, , \quad
   \kappa \dot{\phi}_i \simeq \frac{8}{3} \kappa m_a f_a \left[ \left(
       \frac{\lambda}{3} - 4 \right)
     \frac{\Omega_{\phi0}}{\Omega_{r0}}  a_i \right]^{-2} \, .
 \end{equation}
 
 \subsection{Quadratic potential}
 Not surprisingly, the expressions for the initial SF values are more
 involved in the case of the quadratic potential, because of the
 extreme fine tuning that is required to keep the SF slowly-rolling at
 early times.

 Using the different formulas in Sec.~2 of
 Ref.~\cite{Urena-Lopez2016}, we first write the SF variables in
 terms of the polar ones,
 \begin{equation}
   \label{eq:4}
   \kappa \phi = -2  \Omega^{1/2}_\phi \cos(\theta/2) \, y^{-1}_1 \, ,
   \quad \kappa \dot{\phi} = \sqrt{6} H \Omega^{1/2}_\phi
   \sin(\theta/2) \, .
 \end{equation}
 The initial values of the above quantities are then calculated,
 after some tiresome but otherwise straightforward algebra, from
 the approximated expressions,
 \begin{subequations}
   \label{eq:5}
   \begin{eqnarray}
     \kappa \phi_i &\simeq& - \left[ \frac{36}{25} \left( 1 +
     \frac{36}{\pi^2} \right) \right]^{3/8} \Omega^{-3/8}_{r0} \,
     \Omega^{1/2}_{\phi 0} \, \left( \frac{m_a}{H_0} \right)^{-1/4} \,
     ,   \label{eq:5a} \\
     \kappa \dot{\phi}_i &\simeq& \frac{\sqrt{6}}{5} \left[
     \frac{36}{25} \left( 1 + \frac{36}{\pi^2} \right) \right]^{3/8}
                                  \Omega^{-7/8}_{r0} \,
                                  \Omega^{1/2}_{\phi 0} \, m_a \,
                                  a^2_i \, ,   \label{eq:5b}
   \end{eqnarray}
 \end{subequations}
 where we assumed that $\theta_i \ll 1$, and then $\cos(\theta_i/2)
 \simeq 1$ and $\sin(\theta_i/2) \simeq \theta_i/2$.

As firstly pointed out in Ref.\cite{Mishra:2017ehw}, the initial
value of the SF variable, see Eq.~(\ref{eq:5a}), is proportional to
the SF mass in the form $\kappa \phi_i \sim m^{-1/4}_a$,
independently of the initial value of the scale factor $a_i$. The
coefficient in the aforementioned proportionality is provided by
numerical factors related to the time at which SF oscillations start,
and to the present values of the radiation and SF density
parameters. On the other hand, the initial value of the SF time
derivative, see Eq.~(\ref{eq:5b}), is proportional to $a^2_i$. For the
numerical examples reported in Sec.~\ref{sec:cosmological-se} above
$a_i = 10^{-14}$, and then for all practical purposes one can safely
take $\kappa \dot{\phi}_i =0$.

\section{Initial conditions for linear density
  perturbations  with a cosh potential \label{sec:init-cond-line}}
Here we show that the growing mode of the linear density
perturbations for a SFDM model with a cosh potential, which is the
attractor solution clearly seen in Fig.~\ref{fig:deltas} for large
scales, can be readily obtained from Eqs.~(\ref{eq:dp}).

Firstly, the condition on large scales means that we can neglect all
the scale-dependent terms in the equations of motion, that is, $k^2
/k^2_J \ll 1$. Secondly, we assume the standard growing mode of the
metric perturbation during radiation domination, which is of the form
$\bar{h} = \bar{h}_i (a/a_i)^2$, where $\bar{h}_i$ is a constant
coefficient. Thirdly, we take the attractor values for the background
variables, namely $\Omega_\phi = -12/\lambda$, $\cos \theta = -1/3$
and $y_1 \simeq \sqrt{8}$ (see Eqs.~(\ref{eq:scaling})
and~(\ref{eq:initialb}), respectively). Under the above assumptions,
the equations of motion for the density perturbations, from
Eqs.~(\ref{eq:dp}), explicitly read,
\begin{equation}
  \label{eq:6}
  \delta^\prime_0 = - \sqrt{8} \delta_1 - \frac{4}{3} \bar{h}_i e^{2N}
  \, , \quad \delta^\prime_1 = - \delta_1 + \frac{2}{\sqrt{2}}
  \delta_0 - \frac{\sqrt{8}}{3} \bar{h}_i  e^{2N} \, ,
\end{equation}
where, as before, the number of $e$-folds is given by $N =
\ln(a/a_i)$.

Eqs.~(\ref{eq:6}) can be solved to find the growing
solutions induced upon the SF variables by the metric perturtation
$\bar{h}$, which are: $\delta_0 = -(2/15) \bar{h}$ and $\delta_1 = -(2
\sqrt{8}/15) \bar{h}$. Given that the growing solution for standard
CDM linear pertubations is $\delta_{\rm CDM} = -(1/2) \bar{h}$, those
of SFDM can also be written as: $\delta_0 = (4/15)\delta_{\rm CDM}$
and $\delta_1 = (4\sqrt{8}/15) \delta_{\rm CDM}$.



\bibliographystyle{JHEP}
\bibliography{references}



\end{document}